# Critical Points with High Accuracy and Fluctuation Origin of 2 and 3-Dimensional Ising Models

You-gang Feng *

Department of Basic Sciences, College of Science, Guizhou University, Cai-jia Guan, Guiyang, 550003 China

**Abstract.** We proposed a new universal method for significantly increasing accuracy of critical points of 2 and 3-dimensional Ising models and exploring fluctuation mechanism. The method is based on analysis of block fractals and the renormalization group theory. We discussed hierarchies and rescaling rule of the self similar transformations, and define a fractal dimension of an ordered block, which minimum corresponds to a fixed point of the transformations. By the connectivity we divide the blocks into two types: irreducible and reducible. We find there are two block spin states: single state and $k$-fold state, each of which relates to a system or a subsystem described by a block spin Gaussian model set up by mathematic map. Using the model we obtain a universal formula of critical points by the minimal fractal dimensions. We computed the critical points with high accuracy for three Ising models. It is the first time to find a critical point only requires a fractal edge, which causes fluctuations, and the point acts as a fluctuation attractor. Finally, we discussed a possibility of different block spins at $T_c$.

## 1. Introduction

The first exact result for a 2-dimensional Ising model was obtained by Kramers and Wannier [1]. Onsager derived an explicit expression for the free energy in the absence of an external field and established the precise nature of the specific heat singularity [2]. A transfer matrix method was employed, and in the calculating processes the periodic boundary conditions were used. However, the conditions changed topological property of the original system as mentioned by [3]. These methods are not universal since they cannot be used in 3-dimensional Ising models. Widom pointed out that as the distance from a critical point is varied, the scale of thermodynamic functions can be changed but their function's form [4]; an idea of scaling was proposed. Kadanoff applied the idea to Ising model and in so doing opened the way for the modern theory of critical phenomena set up by Wilson [5-7]. A concept of block spin was introduced: a block spin is an ordered block containing lattice spins and keeping the symmetries of original system. The renormalization group theory indicates the critical point corresponds to a self similar transformation fixed point, and a block spin system has the same thermodynamic properties as the ones of the original system near the critical point. According to the theory of fractals the transformations have fractal structures, it means an ordered block has fractal dimension [8]. As we mentioned in the review [9], an evident deficiency of the renormalization group theory is just that it didn't make any fractal analysis of the blocks. There was blindness in selection of the block sizes. Now that a critical point relates to a unique fixed point of the transformations the block size should relate to it. Unfortunately, we have not seen any relationships between the critical point and the edge in the theory. Although the theory is universal, but the lack of fractal analysis leads to introduction of some approximate methods such as the coarse graining and the decimation, which never give us any critical points with high accuracy. In addition, we notice that although there are some statistical functions of fluctuations such as spin density correlation function and space correlation function, they don't concern internal structures, so the fluctuation mechanism is unknown still. Now that the fluctuations take place at the critical temperature there should be certain relationships among the fluctuations, the critical point and the critical temperature. Studying the fractal structures of blocks may help us explore a universal method, which will significantly increase the accuracy of critical points, set up the relationships and reveal an origin of the fluctuations. This paper aims to find such a universal method. The paper is organized as follows. In Sec. 2, we set up a rescaling rule in the transformations, and defined fractal dimensions for an ordered block (sub-block). In Sec.3 we divide blocks into two types by their connectivity. Giving up Wilson's supposition that a block spin always equals an original lattice spin, we obtain relationships among block spins, fractal dimensions and coordination numbers under the rescaling rule, which change the complicated calculating the block spin interaction into the simplified calculating their fractal dimensions. In Sec.4, using mathematic map we set up a block spin Gaussian model solved accurately, and find the critical point linking to the minimum of block spin and its coordination number. In Sec.5 we analyze topologically blocks and find there are two types of block spin states, which correspond to two independent subsystems for a system. We obtain a formula of a system critical point related to its subsystem critical points. In Sec.6 we proved a unique edge fixed point determines the minimal fractal dimension, by which obtained a universal formula of the critical points. We then calculated the critical points of three systems and find a critical point with a fractal edge acts as an attractor, around which the fluctuations go on from time to time, and in the edge adjustment different block spins will occur at $T_c$. A conclusion is in Sec.7.

---

* *E-mail: ygfeng45@yahoo.com.cn*

## 2. Hierarchies, rescaling rule and fractal dimensions

We pointed out that the block edges should be finite in the self similar transformations, so that there is a unique fixed point of the transformations [9]. At that time the correlation length changes into infinity only by iterations, which show that the transformations are under the necessity of hierarchies. The hierarchies are just the scale invariant leading to a rescaling rule directly: We call an original lattice the zeroth order lattice, some of them construct a first order block by the transformation, and the block is said to be on the first hierarchy. On the ($m+1$)th hierarchy there are only the ($m+1$)th order blocks independent of each other. A ($m+1$)th order block contains the $m$th order lattices, which are just the $m$th order blocks before rescaling. According to the rule after the formation of the ($m+1$)th order blocks the $m$th order blocks should shrink to $m$th order lattices. It is clear that the inside space of the ($m+1$)th order block is just the outside space of the $m$th order lattices, the block space is of dimensions $D$, see Eq.(1). As a lattice the inside of a $m$th order lattice is indistinguishable. Thus the rescaling rule exists only between two neighbor hierarchies. A block spin is $S$ and a lattice spin is $s$, and $s^2 = 1$. From the above analysis we see that on the one hand the renormalization group theory indicates the fixed point of the transformations corresponds to a unique critical point for a given system. On the other hand all of identical ordered blocks can exert the transformations, in which there is a unique fixed point due to the fixed point theory [9]. So there is a unique edge related to the critical point. Since there are interactions of ordered blocks we can't neglect the lattice spins in it, which result in an introduction of a fractal dimension of the ordered block: Let an ordered block edge be $n$, its lattices be covered by open balls with diameter $1/n$ and the ball number be $P$ at least, the fractal dimensions $D$ be defined as [10,11]

$$D = -LnP / Ln(1/n) = LnP / Ln(n) \qquad (1)$$

On the contrary, if a disordered block on a hierarchy is an independent unit not containing any other ordered regions apart from the lattice spins, it is meaningless to define its dimension. Since there are no interactions between disordered blocks or between a disordered block and an ordered block, so the disordered block is equivalent to a region not containing any lattice spins. Such an equivalent description is reasonable, because the disordered block acts as an empty set in the transformations, and an empty set has no dimensions in a mathematic sense. In Eq. (1) the edge values should guarantee the fractal dimensions reasonable, otherwise the relevant fractal structure will not exist. By Eq.(1), the inside space of an ordered block amounts to a super cube of dimensions $D$ with edge $n$ and volume $P = n^D$, in this sense the fractal dimensions are also called capacity dimensions [10,11]. See Fig.1, a triangle with vertices of $P = (n+1)(n+2)/2$ has $n^2$ cells, where a cell is a minimal simplex (a minimal triangle). If we put a spin on each vertex, the triangle then becomes a block spin containing $P$ lattice spins with edge $n$, where we define the distance of two nearest neighbor lattices as a unit length. By Eq. (1), the fractal dimensions $D_{tr}$

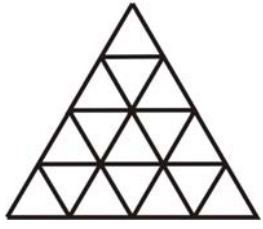

Fig.1 Shown is an irreducible block of the triangle lattice of $n = 4$.

of the ordered block are defined by

$$D_{tr} = \{Ln[(n+1)(n+2)/2]\}/Ln(n) \tag{2}$$

From Eq. (2) we see that $D_{tr}$ is an edge function. For the plane square lattice, if we define directly a fractal dimension of a square block by Eq. (1), it is $D = \{[Ln(n+1)^2]/Ln(n)\} > 2$ larger than its embedding space dimensions. Such a structure will not exist. As a result a square lattice block should be decomposed into $k$ sub-blocks, each of which has a fractal dimension smaller than 2, they are ordered and identical. Let there be $k$ sub-blocks in a square block. For $k=2$ as illustrated in Fig.2, a sub-block has the same edge as the square block edge. By Eq. (1) we get its fractal dimension

$$D_{sq} = \{Ln[(n+1)^2/2]\}/Ln(n) \tag{3}$$

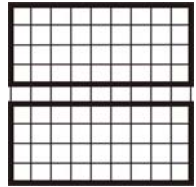

Fig.2 In the plane square lattice system a reducible block can only contain two sub-blocks, which interaction is along the normal direction of the edges.

If $k = 3$ an additional sub-block will intervene between the two sub-blocks, the additional block can't transform in the same way as the other sub-blocks do: If the edges of its neighbor sub-blocks change to semi-infinity, its edges have to keep limit. Such non-uniform transformation will break the original symmetries, which means the case of $k = 3$ is impossible. If $k = 4$, there are four smaller squares, their fractal dimensions are also unreasonable. The cases of $k > 4$ are similar to the case of $k = 3$. So the case of $k = 2$ is unique. Similarly, the fractal dimensions of an ordered sub-block of the cube lattice are

$$D_{cu} = \{Ln[(n+1)^3/4]\}/Ln(n) \tag{4}$$

As illustrated in Fig.3 a cube is subdivided into $k = 4$ cuboids with the same edges as the cube block edge. If *k*=2 a sub-block has a square lattice in a 2-dimensional section, the square lattice dimension is larger than the section dimension (the square lattice has the same edge as the cube edge), it is unreasonable, so $k \neq 2$; so does $k \neq 3$. If $k = 8$ there are 8 cube blocks with smaller edges, such decomposition is inefficient. If $k = 5$, 6, 7 or $k > 8$, the transformations will break the system symmetries. This means the sub-blocks can only be four identical cuboids. The changing of fractal dimensions directly reflects the transformations. Since the fractal dimensions are an edge function the edge fixed point certainly accords with the transformations' one.

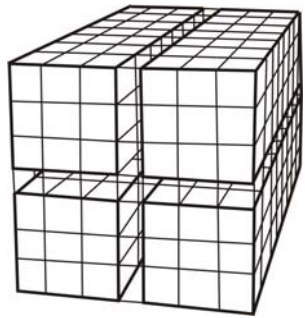

Fig.3 Illustrated is a reducible block in the cube lattice with four sub-blocks and $n = 7$.
The sub-block's interaction is parallel to the normal of their side faces larger area.

## 3. Two types of blocks and systems

An ordered block is simply connected tantamount to a single point space, so it can shrink to a lattice. This characteristic is a mathematic basis for the rescaling rule. Both of triangle and tetrahedroid are simply connected, they are mathematically called simplexes [12]. Square is a mathematic complex, it can be usually decomposed into simplexes, triangles. However, there are no next-nearest neighbor interactions in the model, and a square lattice doesn't contain any triangle lattices. So a square lattice of the model differs from a pure mathematic complex, we cannot decompose it into any triangles as usually. As mentioned in Sec.2 a square block should be decomposed into two sub-blocks. In a square lattice the simplest unit is a square still, it may be regarded as a simplex with a certain physical condition. Thus its sub-block may become simply connected under the condition. The same is true for a sub-block of a cube. We divide the lattice systems into two types: an irreducible system and a reducible system. In an irreducible system a block keeping all the system symmetries is called irreducible block, the triangle block of the triangle system and the tetrahedroid block of the tetrahedroid system are of them. Some identical irreducible blocks can form a new larger irreducible block making the transformations go on. In a reducible system a block keeping all of the system symmetries is called a reducible block, of which a square block and a cube block are. A reducible block should consist of $k$ identical sub-blocks, each of which is ordered and has reasonable fractal dimensions $D$. At that moment as a sum space of its sub-block spaces the disordered reducible block has a non-zero dimension, but the transformation cannot carry out: Although the sub-blocks keep the original symmetries as far as possible, their symmetries are only the partial original ones. For example, in the square lattice system the nearest neighbor interactions of lattice spins are in 2-dimensional directions, but the nearest interactions of the sub-block spins are only in 1-dimensional direction, in the normal direction of the edges, referring Fig.2. If the sub-blocks execute the transformation directly the transformation will break the original symmetries, thus it is impossible. The transformation in a reducible system takes two steps: First, $k$ sub-blocks form in a reducible block. Second, the $k$ sub-blocks make the reducible block ordered by their interactions. The two steps are simultaneous. After that, by the same way some ordered reducible blocks can compose a new bigger reducible block making the transformations on a higher hierarchy. The difference in the fractal structures between the two types of blocks are naturally resulted from their connectivity [9]. According to topology an ordered reducible block is equivalent to a product space of its sub-block spaces [12], it is a simply connected space of dimensions $D^k$. Before shrinkage the ordered reducible blocks can only exist in a space of dimensions $N'$, and $N' \geq D^k$, so we can't illustrate it in the embedding space of the original system. When they shrink to lattices they can exist in the space where the original system lies. When we divide a disordered reducible block into $k$ sub-blocks we consider it as a sum space of its sub-block spaces, a $k$-fold connected space in the embedding space of the original system, and so we can illustrate it as in Figs.2 and 3. We are not concerned about how a disordered reducible block becomes ordered by its sub-block spin interaction, not calculating directly the interaction. We are only concerned about the results such as Eq. (7), (25) and (26) to be used in the partition function.

Now that different sizes accord with different fractal dimensions, for those blocks with different fractal dimensions what spin values they will have? Let a block spin be $S$ and the energy of interaction between two nearest neighbors denoted by $y_1$ and $y_2$ be $JS^2$, where $J$ be a coupling constant. As the transformations a new ordered block formed by the above two blocks on a higher hierarchy, after rescaling the two block spins become two nearest neighbor lattice spins in the new block, they then be denoted by $f(y_1)$ and $f(y_2)$ with each value of $s$, $s^2 = 1$; their interacting energy be $js^2$, where $j$ be a coupling constant in the $D$-dimensional space, which is the inside space of the new block. Let $d[f(y_1), f(y_2)] = js^2$ and $d(y_1, y_2) = JS^2$, the transformation is virtually a contraction map [8], so $d[f(y_1), f(y_2)] = r \cdot d(y_1, y_2)$, $js^2 = rJS^2$, which shows there is a quantitative relation between $js^2$ and $JS^2$. After the transformation and rescaling on the $(m+1)$th hierarchy a block originally called the $m$th order block has changed to a new lattice, a $m$th order lattice. As a lattice it lies the inside of the $(m+1)$th order block, so its outside space is just the inside space of the $(m+1)$th order block of dimensions $D$. For an observed object, whenever it serves as a lattice spin the interacting energy equals $js^2$; as a block spin, however, the energy is $JS^2$. As mentioned in Sec.3 a lattice in a block of dimensions $D$ can be equivalently regarded as a lattice in a super cube of dimensions $D$. It is well known that a coordination number of a $D$-dimensional cube is $2D$, so the total magnitudes of interacting energy of a lattice spin with all its nearest neighbors inside the $(m+1)$th order block equal $2Djs^2$. As a $m$th order block spin before rescaling, however, the total interacting energy of it with all its nearest neighbors, in its outside space, equal $ZJS^2$, where $Z$ is a coordination number of the block spin. In fact, the lattice spin and the block spin are the same observed object described by two previous artificial versions, so these different descriptions must be equivalent in magnitudes, which leads to an equality under the rescaling rule

$$ZJS^2 = 2Djs^2 \qquad (5)$$

Let $K = J/(k_B T)$, $k_B$ the Boltzman constant, $T$ temperature, and Eq. (5) becomes

$$ZKS^2 = 2Djs^2/(k_B T) \qquad (6)$$

Where $Z$ is a constant for a given system, $s^2 = 1$, and $j$ is the coupling constant in the space of dimensions $D$, so $j$ relates to $D$, which means $K$ and $S^2$ also relate to $D$, thus they are not independent of each other. In the renormalization group theory [6], the Author preferred to suppose $S^2 = s^2$, namely, the block spins always are constant, so that $K$ is only determined by $D$ at $T_c$ due to Eq.(6). Since the fractal dimension is an edge function, $K$ is certainly determined by the edges. In fact, $K_c$ is merely governed by an edge fixed point, which will be proved in Sec.6. We think there is not enough evidence to guarantee the supposition of $S^2 = s^2 = 1$ to be correct for any systems and blocks no matter what sizes and shapes the blocks will have, and what type of systems such as the triangle lattice, the square lattice or the cube lattice the blocks lay. The coupling of block spins virtually is lattice spins' one shown in Eq. (5). Thus we have sufficient reasons to suppose that $J = j$ and $S^2 \neq s^2 = 1$, so Eq. (5) becomes

$$ZS^2 = 2D \qquad (7)$$

Eq. (7) indicates that differing from the lattice spins, which are always constant, the block spins can change with their fractal dimensions. In such a case the $S^4$-model will not be applicable [7]. Here we emphasize that we derived Eq. (5), (6) and (7) by the same observed object under the same rescaling rule, so the $J$ and $j$ in Eq. (5) are only fit for the same system. Each subsystem of a reducible system (see Sec.5) has its own equalities similar to Eq. (5) and (7). The applicability of Eq. (6) and (7) will be discussed further in Sec.5. For a given system the coordination number is a constant, so the values of block spins are only determined by the edges due to Eq. (1) and (7). Our supposition that the block spins can change gives us a chance to set up a block spin Gaussian model solved accurately.

## 4. Block spin Gaussian model

### 4.1. Partition function and free energy

On the one hand all of ordered blocks with finite edges can infinitely exert the transformations, on the other hand the transformation fixed point is unique. So we infer that there is a special edge determining the fixed point. When we approach a critical temperature the correlation length becomes larger, but always limited, which may lead to a finite transformation hierarchy. Thus we may consider the block spins as independent variables provided that the correlation length is not larger than a block' size. According to Ergodic hypothesis there are ordered blocks and disordered blocks, with a variety of shapes and sizes before the critical temperature in thermodynamic equilibrium. For example in Fig.4, three triangles are the nearest neighbor ordered blocks, among them the blank space represents a disordered block, which size is smaller than an ordered block's. There are two cases: In the first case there are infinite

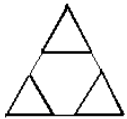

Fig4. Illustrated is a disordered block among three ordered blocks identical ordered blocks with one size and infinite identical disordered blocks with another size simultaneously, and they keep the original symmetries, as illustrated in Fig.4. In such transformations a disordered block is tantamount to an empty set, which spin value and dimension are zero. In the second case there are only infinite identical ordered blocks, as in the square lattice system, and the system can make the transformations infinitely. In other cases apart from the above two cases there are no infinite transformations making the system ordered. For the studying of critical phenomena, we only need to consider the cases of infinite transformations. By mapping we now set up a new system with infinite lattices: the directions of the new lattice spins are the same as the block spins', the absolute values of all new lattice spins are equal. The new system itself cannot execute any transformations. First, the new system keeps all symmetries of the original system. Second, as the block can change no matter what sizes a block will have a new lattice always corresponds to the block. This supposition means a new lattice spin can change and its spin values involve all possible values of the block spins. In addition, we allow the new lattice spin value to equal zero, which also is the statistical average value of block spins $S = <S> = 0$ before the critical temperature. For the original system this is a uniform disordered state, in which there are only identical disordered blocks keeping the original symmetries. Such a state possibly occurs as a result of the Ergodic hypothesis. Last but not least, no matter what sizes a block will have, we always consider the distance between two nearest neighbors a constant, which is just the distance of adjacent new lattice spins. It is obvious that the new system has the same critical point as the original. It is also clear when we discuss the transformations we analyze the original system, when we calculate the critical point we use the new system. Our calculation is similar to [13-15], in the process we mention "lattice" or "lattice spin" we mean the lattice or the lattice spin of the new system.

Suppose that on the $m$ th hierarchy the correlation length is less than a block size, each block spin is regarded as an independent variable, so the statistical laws can be used to describe the new system. The space of interaction of two adjacent lattice spins is as the same space of the original system, the Euclidean space of dimensions $N$. In the new system of the triangle lattice $N = 2$, the dimensions of their lattice vector space and reciprocal lattice vector space also are $N = 2$, respectively. The methods of calculating bases of lattice vector space, bases of reciprocal lattice vector space and Brillouin zones are commonly applied in solid state physics [16, 17]. In Fig.1 we select a lattice position as an origin of a 2-dimensional Cartesian coordinate system (the new system and the original system has the same shape and symmetries, but their lattices have the different meanings). We set up a solid state physical cell, which bases are $\boldsymbol{a}_1 = a\boldsymbol{i}$, $\boldsymbol{a}_2 = (a/2)[\boldsymbol{i} + \sqrt{3}\boldsymbol{j}]$, where $a$ is a cell edge, which also is a lattice constant, $\boldsymbol{i}$ and $\boldsymbol{j}$ are unit vectors for $x$-axis and $y$-axis, respectively. The bases of the reciprocal lattice vector space are given by $\boldsymbol{b}_1 = (2\pi/a)[\boldsymbol{i} - (1/\sqrt{3})\boldsymbol{j}]$, $\boldsymbol{b}_2 = (2\pi/a)(2/\sqrt{3})\boldsymbol{j}$. The number of the reciprocal lattices with the same distance $(4\pi/a)(1/\sqrt{3})$ from the origin is 6, which is just the coordination number, and their coordinates are $[\pm 2\pi/a, \mp 2\pi/(\sqrt{3}a)]$, $[0, \pm 4\pi/(\sqrt{3}a)]$ and $[\pm 2\pi/a, \pm 2\pi/(\sqrt{3}a)]$. The first Brillouin zone is a hexagon, over which the values of the bases $q_x$ and $q_y$ of the reciprocal lattice vector space go from the negative to the positive are $-4\pi/(3a) \leq q_x < 4\pi/(3a)$ and $-2\pi/(\sqrt{3}a) \leq q_y < 2\pi/(\sqrt{3}a)$. At first we consider that the values of block spins don't change, so do the new lattice spins. In the absence of an external field the Hamiltonian of the new system takes

$$H^* = K\sum_{(i,j)} S_i S_j , (S_i, S_j = \pm S_{tr}) \tag{8}$$

Where $K \equiv J_{tr}/(k_B T)$, $J_{tr} = j$ due to our supposition, $\sum_{(i,j)}$ is sum over all nearest neighbors and $S_{tr}$ is a constant temporarily. However in fact, on the one hand by Eq. (7) and (2) the coordination number is constant for a given system, so only the edges determine $S_i$ and $S_j$. On the other hand the edges can change and their fluctuations should obey the central limit theorem in thermodynamic equilibrium [15, 18]. Thus, there are fluctuations of the new lattice spins and they can be described by Gaussian distribution as the central limit theorem:

$$W = \prod_{j=1}^{N_c} \exp[-S_j^2/(2\langle S_{tr}^2 \rangle)] \tag{9}$$

Where $N_c$ is the total number of cells, $N_c \to +\infty$ and $\langle S_{tr}^2 \rangle$ is a mean square of the new lattice spins. In the renormalization group theory under the supposition that a block spin always equals a lattice spin the $S^4$-model is a preferential model compared with the Gaussian model [7]. However, when we consider that the block spins depend on the edges and can change the block spin Gaussian model is more excellent than the $S^4$-model. The new system is called a block spin Gaussian model for Eq. (9), which may be considered as either an approximate supposition or a realistic law. Combining Eq. (8) and (9) we then get the partition function of the new lattice system

$$Q = \int_{-\infty}^{+\infty} \cdots \int_{-\infty}^{+\infty} \prod_{j=1}^{N_c} dS_j \exp\{K\sum_{(i,j)} S_i S_j - (1/2 < S_{tr}^2 >)\sum_{j=1}^{N_c} S_j^2\} \tag{10}$$

In Eq. (10) for simplicity, we extend the range of the magnitudes of $S_{tr}$ to infinity: $-\infty < S_{tr} < +\infty$ and make the values continuous. Such a procedure is based on that we only focus on a singularity of the free energy rather than on the function values. The later results will reveal that the meaningful values of $S_{tr}$ are absolutely governed by Eq. (7) and (2). The Gaussian weighting factor gives the largest weight to the value $S_j = 0$, the statistical average value of the new lattice spins before the critical temperature. It is true since the system is disordered still before $T_c$. The integration is only carried out in the inscribed circle of the first Brillouin zone. As mentioned by [14, 15] and seen follows, such a calculation is still valid because a contribution to the singularity of the free energy uniquely comes from the range near the origin covered by the inscribed circle. In other words, the singularity only depends on the long-wavelength part of the wave vectors. In Eq. (10) we let the effective Hamiltonian of the new lattice system be $H = K \sum_{(i,J)} S_i S_j - (1/2 < S_{tr}^2 >) \sum_{j=1}^{N_c} S_j^2$. In follows, we introduce Fourier transformations of the spins,

$$S_i = (1/\Omega) \sum_q S_q \exp(i\, q \cdot r_i), \quad S_q = V \sum_{i=1}^{N_c} S_i \exp(-i\, q \cdot r_i), \quad (11)$$

Where $V$ is a cell volume and $\Omega = N_c V$ is the total volume of the system, $N_c$ the total number of cells and the sum over $q$ is restricted in the first Brillouin zone. Let

$$K(r_i - r_j) = K \quad r_i, r_j \text{ nearest} \quad (12.1)$$

$$K(r_i - r_j) = 0 \quad \text{others} \quad (12.2)$$

Using Eq. (11) and (12) we make the effective Hamiltonian become

$$H = (1/2)\{[\sum_{i,j} K(r_i - r_j) S_i S_j] - [(1/<S_{tr}^2>) \sum_{i,j} \delta(r_i - r_j) S_i S_j]\} \quad (13)$$

Where the two terms $\sum_{i,j}$ are, different from $\sum_{(i,j)}$ in Eq. (8) and (10), independent sums over $i$ and $j$, respectively, and

$$\delta(r_i - r_j) = 1, r_i = r_j \quad (14.1)$$

$$\delta(r_i - r_j) = 0, \text{ others} \quad (14.2)$$

Using Eq. (11), we get a Fourier transformation of the form

$$\sum_{i,j} f(r_i - r_j) S_i S_j$$

$$= (1/\Omega V N_c) \sum_{q_1, q_2} S_{q_1} S_{q_2} \sum_i \exp[i(q_1 + q_2) \cdot r_i] \sum_j f(r_i - r_j) \exp[-i q_2 \cdot (r_i - r_j)] \quad (15)$$

Let

$$f(q_2) = \sum_j f(r_i - r_j) \exp[-i q_2 \cdot (r_i - r_j)] \quad (16)$$

Where $f(q)$ and a Fourier component of $f(r)$ are a constant apart at most. Noticing the orthonormality:

$$(1/N_c) \sum_i \exp[i(q_1 + q_2) \cdot r_i] = \delta(q_1 + q_2) = 1, q_1 + q_2 = 0 \quad (17.1)$$

$$= 0, \text{ others} \quad (17.2)$$

Where $q_1$ and $q_2$ take all possible values and renew: $q_2 = q$, $q_1 = -q$. $S_i$ are real numbers, $S_q^* = S_{-q}$, where $S_q^*$ is conjugate to $S_q$. Eq. (15) becomes

$$\sum_{i,j} f(r_i - r_j) S_i S_j = (1/\Omega V) \sum_q f(q) |S_q|^2 \qquad (18\text{-}1)$$

Analogously

$$\sum_{i,j} K(r_i - r_j) S_i S_j = (1/\Omega V) \sum_q K(q) |S_q|^2 \qquad (18\text{-}2)$$

From Eqs.(16) and (12), we get

$$K(q) = \sum_j K(r_i - r_j) \exp[-i q \cdot (r_i - r_j)] = K \sum_\delta \exp(-i q \cdot \delta_{ij}) \qquad (19)$$

where $\delta_{ij}$ is a vector from the lattice $i$ to its nearest neighbor lattice $j$. Eq. (19) is ready to calculate the critical point, see Eq. (22). Inserting Eq. (18) and (19) to Eq. (13), noticing that whenever $f(r_i - r_j) = \delta(r_i - r_j)$, $f(q) = 1$. We rewrite Eq. (13) as

$$H = (-1/2\Omega V) \sum_q [1/<S_{tr}^2> - K(q)] |S_q|^2 \text{ and Eq. (10) as } Q = \int_{-\infty}^{+\infty} \ldots \int_{-\infty}^{+\infty} \prod_q dS_q \exp H, \text{ a}$$

typical Gaussian integration. We then get the partition function of the new system:

$$Q = \prod_q \{(2\pi\Omega V)/[1/<S_{tr}^2> - K(q)]\}^{1/2} \qquad (20)$$

Finally, the new system's free energy takes the form as

$$F = -k_B T Ln Q = (1/2) k_B T \sum_q Ln[1/<S_{tr}^2> - K(q)] + T \cdot const. \qquad (21)$$

**4.2. Determination of a critical point**

From Eq. (21) the singularity of free energy turns up when $K(q)$ equals $1/<S_{tr}^2>$, which corresponds to the critical point. Whenever the temperature $T$ is higher than $T_c$ $K(q)$ always is less than $1/\langle S_{tr}^2 \rangle$, so the approach $K(q) \to 1/\langle S_{tr}^2 \rangle$ results in the maximum of $K(q)$ at $T_c$; inversely, the minimum of $\langle S_{tr}^2 \rangle$. It should be emphasized this is directly a result of the singularity and is regarded as a mutation of statistical regular pattern rather than a statistical mean value at $T_c$. As the critical point is unique the maximum of $K(q)$ should be unique, too. Only $S_{tr,\min}^2$ can do this for $S_{tr,\min}^2$ is a unique minimum, see Sec.6. So the two extreme values are one-to-one. The coordinates of six lattice vectors associated with $\delta_{ij}$ near the origin are $(\pm a, 0)$, $(\pm a/2, a\sqrt{3}/2)$ and $(\pm a/2, -a\sqrt{3}/2)$. Inserting these to Eq. (19), finding that $K(q)$ reaches its maximum at $q=0$, so $K(0) = 6K_c = 1/S_{tr,\min}^2$. Thus we get

$$K_c = (6S_{tr,\min}^2)^{-1} \qquad (22)$$

The whole deriving process reveals the reciprocal lattice vectors always vanish at the $K_c$. The number 6 in Eq. (22) is just the coordinate number of a block. With the same reason, whenever we know a coordination number $Z$ of an ordered block and its value of $S_{\min}^2$, we may get the system's critical point:

$$K_c = (ZS_{\min}^2)^{-1} \qquad (23)$$

## 5. Two types of block spin states

An ordered reducible block is a result of its sub-block spin interactions, so the block spin state differs from its sub-block spin state. We call a sub-block spin state a single state of spin; there are $k$ single states in a reducible block consisting of $k$ sub-blocks. We call the block spin state a $k$-fold state since the ordered block is a product space of its $k$ sub-block spaces. Thus there are two types of block spin states in a reducible system (in an irreducible system there only is a single state). Each sub-block spin preserves its own independence and is simply connected. Before ordered the reducible cannot be contractible, it is a sum space of its sub-block spaces as illustrated in Figs.2 and 3. The correlation length is not greater than a sub-block size. For example see Fiq.2, in the square lattice system, $k = 2$ there are two sub-blocks. As a single state a sub-block spin is $S_{11}$, its coordination number is $Z_{11} = 2$. When the reducible block becomes ordered, for the moment it acts as a simply connected space; the correlation range is as large as its size and it is a product space of its sub-block spaces. The ordered reducible block is of dimensions $D^k$, where $D$ is the dimensions of a sub-block. At the moment it will be able to shrink, so its coordination number is just the original one of the system. For example, in the square lattice system a reducible block spin $S_{12}$ is in a double state, its coordination number $Z_{12} = 4$. Thus, on the $m$th hierarchy there exist simultaneously the $m$th order sub-block spins and the $m$th order reducible block spins. The two types should correspond to two statistically independent subsystems. Moreover, as illustrated in Fig.2 the sub-block spins interact only along the normal directions of the block edges, these interacting sub-block spins make up a subsubsystem. The first subsystem consists of $N_{11}$ subsubsystems, and $N_{11} \to \infty$. These subsubsystems are geometrically parallel to each other. There are no the nearest neighbor interactions of the sub-block spins between subsubsystems, so all subsubsystems are independent of each other, and they are identical. Each subsubsystem can be described by a 1-dimensional block spin Gaussian model since the sub-block spins interact only along 1-dimensional direction. Thus, the partition function $Q_{11}$ of the first subsystem can be written as $Q_{11} = N_{11}Q_{11ss}$, where $Q_{11ss}$ is the partition function of a subsubsystem. Obviously, all of the subsubsystem partition functions have the same form and a common critical point $K_{c1}$, which is just the critical point of the first subsystem. The partition function of the second subsystem is denoted as $Q_{12}$. Thus the partition function $Q_{sq}$ of the square lattice system is written as $Q_{sq} = Q_{11}Q_{12} = N_{11}Q_{11ss}Q_{12}$, where the product form of $Q_{11}$ and $Q_{12}$ implies there exist two statistically independent subsystems simultaneously. A discussion analogous to the irreducible system in Sec.4 indicates the critical point of the system only depends on $Q_{11ss}$ and $Q_{12}$. We get a logarithmic form of the partition functions related to the free energy $LnQ_{sq} = LnN_{11} + LnQ_{11ss} + LnQ_{12}$. We see that the singularity of the free energy is determined by the singularities of both $LnQ_{11ss}$ and $LnQ_{12}$, and the term $LnN_{11}$ has no contributions to the critical point since it doesn't contain any spins. Let the system's critical point be $K_c$ related to the singularity of $LnQ_{sq}$, the critical point $K_{c1}$ to the singularity of $LnQ_{11ss}$, the critical point $K_{c2}$ to the singularity of $LnQ_{12}$; $K_c$ should be the sum of $K_{c1}$ and $K_{c2}$, so

$$K_c = K_{c1} + K_{c2} \tag{24}$$

A similar way can be used to treat the cube lattice system. Thus, we can obtain a critical point of a reducible system such as the square lattice system by computing the critical points of its subsystems.

In order to calculate the two critical points we should obtain relationships among block spins, fractal dimensions and coordination numbers under the rescaling rule. On the *m*th hierarchy the *m*th order sub-blocks and the *m*th order ordered reducible blocks exist simultaneously and they don't shrink to lattices temporarily on the hierarchy due to the rescaling rule. But on the next hierarchy, on the (*m*+1)th hierarchy, in the first subsystem of the square lattice system a $m$ th order sub-block spin $S_{11}$ with coordination number $Z_{11}$ and coupling constant $J_{11}$ has become a (*m*+1)th order lattice spin $s$ with coupling constant $j_{11}$ in a space of dimensions $D_{sq}$ after rescaling, the space is just the inside space of the $(m+1)$ th order sub-block equivalent to a super cube of edge $n$, volume $n^{D_{sq}}$ and coordination number $2D_{sq}$. Meanwhile in the second subsystem a $m$th order reducible block spin $S_{12}$ with coordination number $Z_{12}$ and coupling constant $J_{12}$ has become a ($m$+1)th order lattice spin $s$ with coupling constant $j_{12}$ in a $(m+1)$th order ordered reducible block after rescaling, the inside space of the $(m+1)$th order ordered reducible block is of dimensions $D_{sq}^2$, with coordination number $2D_{sq}^2$. Thus on the ($m$+1)th hierarchy, using Eq. (7) we get

$$Z_{11}S_{11}^2 = 2D_{sq}, \qquad Z_{12}S_{12}^2 = 2D_{sq}^2 \qquad (25)$$

Eq. (25) implies our suppositions of $J_{11} = j_{11}$ and $J_{12} = j_{12}$, but we can't infer that the equalities $J_{11} = j_{12}$, $J_{12} = j_{11}$ or $J_{11} = J_{12}$ since these constants link to two different subsystems, they are incomparable with each other (see the end of Sec.3). Fig.2 only illustrates a disordered reducible block as a sum space of two sub-block spaces. But we can't illustrate an ordered reducible block as a product space of the two sub-block spaces in a 3-dimensional space since $D_{sq}^2$ >3. With the same reason, in the cube lattice system we get

$$Z_{21}S_{21}^2 = 2D_{cu}, \qquad Z_{22}S_{22}^2 = 2D_{cu}^4 \qquad (26)$$

Where $S_{21}$ represents a ($m$+1)th order sub-block spin (the single state) with coupling constant $J_{21}$ and coordination number $Z_{21}$, $D_{cu}$ determined by Eq.(4). The $S_{22}$ represents the ($m$+1)th order reducible block spin (the fourfold state) with coupling constant $J_{22}$ and coordination number $Z_{22}$. The inside space of the ($m$+1)th order ordered reducible block is of dimensions $D_{cu}^4$, which is a product space of its four the ($m$+1)th order sub-block spaces, each of which is of dimensions $D_{cu}$. Both $J_{21}$ and $J_{22}$ are incomparable as they belong to different subsystems. Seeing Fig.3, since there are only the nearest neighbor interactions in the Ising model, for the cube lattice system the interactions of sub-blocks are only parallel to the normal of their side faces with larger area, these interacting sub-blocks make up a subsubsystem. The first subsystem consists of $N_{22}$ identical subsubsystems and $N_{22} \to \infty$, each of which can be described by a 2-dimensional block spin Gaussian model since the interactions only occur in a 2-dimensional space. All subsubsystems are geometrically parallel to each other and they don't interact because there are no the nearest neighbor interactions of the sub-block spins between them. These cases are like the cases in the square lattice system, so Eq. (24) also is suitable to the cube lattice system.

## 6. Results and discussion

In order to compute the minimal fractal dimension we consider the edge as a continuous parameter temporarily. For the triangle lattice system by Eq. (2), when the fractal dimensions take their minimum the zero value of the derivative of $D_{tr}$ with respect to $n$ leads to a fixed point equation

$$[(n+1)(n+2)/(2n+3)] \cdot Ln[(n+1)(n+2)/2]/Ln(n) = n \tag{27}$$

The Eq. (27) has a unique edge fixed point $n^*$ due to the Banach fixed point theorem [12, 19]. Computing Eq. (27) and (2) yields a unique minimal dimension:

$$n^* = 14.4955, \qquad D_{tr,\min} = 1.814055098 \tag{28}$$

With the same reason, computing Eq. (3) we get a unique minimum $D_{sq,\min}$ with a unique edge fixed point $n^*$ for the sub-block of the square lattice:

$$n^* = 7.839995, \qquad D_{sq,\min} = 1.779990992 \tag{29}$$

Similarly, by Eq. (4) for the sub-block of the cube lattice a unique minimum $D_{cu,\min}$ with a unique edge fixed point $n^*$ is

$$n^* = 4.749100, \qquad D_{cu,\min} = 2.478143004 \tag{30}$$

All of the fixed points are fractal edges. Combining Eq. (7) with Eq. (23)-(26), we know a unique minimum $D_{\min}$ determines a unique minimum $S^2_{\min}$ and get $K_{c1} = 1/2D_{\min}$ and $K_{c2} = 1/2D^k_{\min}$. We then obtain a universal formula of the critical points related to $D_{\min}$

$$K_c = 1/2D_{\min} + 1/2D^k_{\min} \tag{31}$$

Where the first term on the right side of the above equation relates to a single state; the second does to a $k$-fold state, which will vanish for an irreducible system. It is the first time to find that a critical point only requires a fractal edge, which will result in edge fluctuations at $T_c$. Using Eq. (31) we can numerically compute the critical points. Inserting Eq. (28) in Eq. (31) and noticing the second term vanishes, we obtain $K_c = 0.2756$. Kramers and Wannier got a result [1], 0.2747.  For the square lattice $k = 2$, using Eq. (29) and (31) we get $K_c = 0.4387$. Onsager obtained a result [2], 0.4407. For the cube lattice $k = 4$, by Eq. (30) and (31) we have $K_c = 0.2150$. The Authors of [20] got its series solution of 0.2217, which also is a result of Monte Carlo simulation [21]. All of results don't relate to any fractal edges apart from ours.
Although our supposition that there are blocks with unique size and shape at a critical point helps us get a result with high accuracy, it doesn't eliminate a possibility of existence of different block spins at $T_c$ in the original system. The fractal edges $n^*$ only associate with the critical points. Computing Eq. (2)-(4), we investigated that around a edge fixed point $n^*$ those obviously different edges, $n > n^*$ or $n < n^*$, correspond to the almost same fractal dimensions as the minimums, e.g., in the triangle lattice system: $n_1 = 14$, $D_1 = 1.814091603$; $n_2 = 15$, $D_2 = 1.814092989$; $D_1$ and $D_2$ are very close to $D_{tr,\min}$, see Eq.(28). It is well known the critical fluctuations are resulted from the adjustment of the inner structures. For the self similar transformations the adjustment is just the block edge adjustment. At $T_c$ we see that on one hand the transformations only allow the blocks to take integer edges; on the other hand a critical point just requires a fractal edge. Such a contradiction in the edges forces a system to adjust the edges continuously in order to reach the critical point further. This fluctuation mechanism means the critical point and the critical temperature are different parameters, although the former only appears at the latter. A system can reach closely to a critical point at $T_c$, but will never arrive at the point because the transformations forbid any fractal edges. Meanwhile, the critical point acts as an attractor, around which the edge adjustment goes on from time to time. In the process different block spins will occur.

## 7. Conclusion

The block spin Gaussian model helps us get the critical points with high accuracy and make the calculation simplified. A fractal edge is a fluctuation origin, and in the edge adjustment different block spins occur at $T_c$. This method provides us theoretical basis for the further exploration of critical phenomena in microscopic structures.